\documentclass[12pt]{article}
\usepackage{amsmath}
\usepackage{amssymb}
\usepackage{graphicx}
\usepackage{cite}

\numberwithin{equation}{section}

\begin{document}
\baselineskip=18pt
\begin{titlepage}
\begin{flushright}
{\small KYUSHU-HET-106}
\end{flushright}
\begin{center}
\vspace*{11mm}

{\large\bf%
Probing flavor structure in unified theory with scalar spectroscopy 
}\vspace*{8mm}

Kenzo Inoue, Kentaro Kojima, and Koichi Yoshioka%
\vspace*{1mm}

{\it Department of Physics, Kyushu University, Fukuoka 812-8581, Japan}

\vspace*{3mm}

{\small (March, 2007)}
\end{center}
\vspace*{5mm}

\begin{abstract}\noindent%
The flavor structure in unified theory is probed with superparticle
mass spectrum observed in future particle experiments. A key
ingredient is the generation dependence of scalar mass
non-degeneracy. The observed non-degeneracy in low-energy regime is
shown to provide a direct imprint of flavor structure in high-energy
fundamental theory. The implication from flavor-violating rare process 
is also discussed.
\end{abstract}
\end{titlepage}
\newpage

\section{Introduction}
\label{sec:int}
Supersymmetry is one of the most prominent solution to the hierarchy
problem. A direct consequence of supersymmetry is the existence of
superpartners of the standard model (SM) particles. In addition,
supersymmetry is strongly favored in view of grand unified theory
(GUT). The minimal supersymmetric extension of the SM (MSSM) with
TeV-scale superparticles leads to precise unification of the SM gauge
couplings at a single scale $\sim 10^{16}$~GeV~\cite{GUT}. Thus
supersymmetric GUT is well motivated and has been widely investigated.

The mass spectrum of superparticles, that is soft
supersymmetry-breaking parameters, is determined by
supersymmetry-breaking dynamics and mediation mechanism\cite{mssm}. If
the mediation scale is higher than the unification scale, e.g.\ a
mediation by Planck scale physics, the superparticle mass spectrum is
expected to have some imprints of fundamental theory. This implies
that high-energy physics can be probed through the observed
superparticle spectrum in low-energy regime. Along this line, 
possible implications from physics around the GUT scale have been
investigated in the literature~\cite{PP,D,gutph}.

In this paper we focus on the matter generation structure in unified
theory and show that it can be probed with superparticle mass
spectrum, once observed in near future particle experiments. In GUT
models, quarks and leptons are generally unified into GUT gauge
multiplets. It is thus an interesting subject to study the recently
revealed disparate flavor structure of quarks and leptons.

The MSSM scalar mass spectrum somehow inherits the matter flavor
structure in high-energy theory. In particular, the first and second
generation scalars generally have non-degenerate masses even when 
the universal supersymmetry breaking is assumed at a high-energy scale,
as we will discuss below. We point out that particular combinations 
of MSSM scalar masses are useful for directly extracting the matter
flavor structure. With these combinations, the origin of generation
structure in unified theory is probed through scalar mass spectrum. 
We also discuss the implications and constraints on non-universal
flavor origins in the lepton sector.

The paper is organized as follows. In Section~\ref{sec:gut} the
generation dependent angles are introduced to describe the flavor
dependence of matter structure in unified theory. We examine the
resultant non-universality of scalar mass spectrum. Several scalar
mass relations for extracting these angles are presented. In
Section~\ref{sec:low}, we show with these relations that the flavor
structure is probed from the scalar mass spectrum, which is expected
to be observed in near future particle experiments. Physical
implications to flavor origins are discussed in
Section~\ref{sec:lfv}. Finally we summarize the discussion in
Section~\ref{sec:sum}.

\bigskip
\section{Matter embedding into unified theory}
\label{sec:gut}
\subsection{Generation redundancy}
In grand unified models, the SM matter multiplets are unified into
larger GUT gauge multiplets. In the minimal $SU(5)$ model~\cite{GG},
the electroweak doublet quarks $Q_i$, up-type quarks $\bar u_i$ and
charged leptons $\bar e_i$ are unified into decuplets, and the doublet
leptons $L_i$ and down-type quarks $\bar d_i$ are combined in anti
quintuplets ($i=1,2,3$ denotes the generations). Further in models
with $SO(10)$ or larger unified symmetry, all matter multiplets in one
generation are unified into a single gauge multiplet. The simple and
naive assumption of matter unification is that three-generation matter
multiplets are embedded into GUT multiplets in a completely parallel
fashion.

While the unification of MSSM gauge couplings suggests the existence
of grand unification at high-energy regime, the matter unification
does not seem to be simply realized and the observed quark and lepton
masses and mixing indeed have diverse structures. In particular, the
recently revealed lepton flavor mixing is quite different from that in
the quark sector~\cite{neu}. To realize such diversity is therefore an
interesting subject in unified theory. An available approach to this
subject is to adopt flavor assignment of matter multiplets in
generation-dependent manner. A non-universal matter embedding
naturally emerges if the unified gauge group $G_U$ is enough large to
have the freedom to choose the ways of embedding. For example, in
$E_6$ unified models, a vectorial representation ($27$-plet) doubly
contains the electroweak doublet leptons (and also the down-type
quarks) which cannot be distinguished with respect to the SM
charges. Generally, each low-energy SM matter field $\phi$ could
appear as a linear combination of some distinct fields. In this work,
we consider the situation
\begin{eqnarray}
  \label{twistgen}
  \phi &=& \phi'\cos\theta_\phi +\phi''\sin\theta_\phi,
\end{eqnarray}
where the SM gauge charges of $\phi'$ and $\phi''$ are identical 
to $\phi$, and these two fields are distinctive to each other with
respect to $G_U/G_{\rm SM}$ charges. In this work, the 
parameter $\theta_\phi$ is referred to as the twisting angle 
for $\phi$. The flavor embedding structure is thus parametrized by
twisting angles $\theta_{\phi_i}$ ($i=1,2,3$) in addition to 
the $G_U/G_{\rm SM}$ gauge charges of $\phi'$ and $\phi''$.

A trivial example is $G_U=SO(10)$ with only three $16$-plets
introduced as matter multiplets. That results in all twisting angles
being equal. With extensions of this example, e.g.\ including
additional matter 10-plets of $SO(10)$, the embedding of $Q_i$, $u_i$
and $e_i$ is unique for each generation and 
then $\theta_{Q_i}=\theta_{\bar u_i}=\theta_{\bar e_i}$. On the other
hand, both 16 and 10 plets contain $\bar 5$ multiplets of $SU(5)$
subgroup, and the model contains two candidates for down-type quarks
and doublet leptons in low-energy regime, while these candidates have
different high-energy gauge charges under 
$U(1)\subset SO(10)/SU(5)$. In this case it is therefore natural for
the twisting angles to depend on the 
generations; $\theta_{L_i}\neq\theta_{L_j}$ and 
$\theta_{\bar d_i}\neq\theta_{\bar d_j}$.\footnote{In this work we do
not consider the gauged $B-L$ symmetry which is viable only for
$\theta_i=0$ or $\pi/2$. The continuous values of twisting angles
implies that the potential terms should possess an abelian global
symmetry to avoid rapid nucleon decay.}

In the following we discuss the explicit example of $G_U=E_6$
case~\cite{E6} with three $27$-plets as matter multiplets. The SM
matter embedding into $E_6$ $27$-plets and the notation are summarized
in Appendix~\ref{sec:27}. Our analysis is general and can be easily
extended to other scenarios as long as the
parametrization \eqref{twistgen} is valid. The only assumption we
adopt in this paper is that the gauge coupling unification with the
desert is respected. That is, $SU(5)$ or larger simple group which
contains $SU(5)$ as a subgroup is broken down to $G_{\rm SM}$ at the
unification scale. Then the down-type quarks and doublet leptons are
expressed as follows:
\begin{eqnarray}\notag
  \bar d_i &=& \bar 3^b_i\cos\theta_i+\bar 3^c_i\sin\theta_i, \\
  \label{twist1}
  L_i &=& 2^a_i\cos\theta_i +2^c_i\sin\theta_i,
\end{eqnarray}
where the common angles $\theta_i\equiv\theta_{\bar d_i}=\theta_{L_i}$
is a result of $SU(5)$ gauge symmetry. The multiplets $2^{a,c}$ and
$\bar 3^{b,c}$ are contained in a single 27-plet and characterized by
$U(1)\times U(1)'\subset E_6/G_{\rm SM}$ gauge 
charges [e.g.\ see~\eqref{charges} in Appendix~\ref{sec:27}]. The
other assignments of SM matter fields are uniquely determined
\begin{eqnarray}\notag
  Q_i &=& (3,2)_i, \\
  \bar u_i &=& \bar 3^a_i, \label{twist2} \\ \notag
  \bar e_i &=& 1^a_i.
\end{eqnarray}
Therefore the twisting angles for these matter fields take the
universal, flavor-independent values. Finally we comment that 
non-parallel matter embedding discussed above is deeply suggested by
the recent experimental results of neutrino flavor mixing in the
context of grand unification~\cite{twist} and is motivated by
low-energy observations.

\subsection{Sum rules for induced scalar mass}
With the generation redundancy in unified theory discussed above, 
matter fields with the same SM charges generally have flavor-dependent
$G_U/G_{\rm SM}$ quantum numbers. The corresponding flavor-violating
gauge interactions leave their imprints in scalar mass spectrum in
low-energy effective theory. That is determined by non-vanishing
$D$-term contribution~\cite{D} and renormalization-group (RG)
effects~\cite{PP} in high-energy theory. According to the generation
structure, these effects induce specific types of non-universality for
MSSM scalar mass spectrum. A notable future is that the induced
non-universality is connected to the generation twisting angles
discussed in the previous section, since the flavor-dependent
$G_U/G_{\rm SM}$ charges of matter fields are parametrized by these
angles. Using this fact, one can directly probe into the generation
structure in high-energy theory from observed scalar mass spectrum at
low-energy regime.

Here we study an example of $E_6$ unified theory described 
by \eqref{twist1} and \eqref{twist2}, and examine the resultant scalar
spectrum with nonzero twisting angles. The following discussion is
similarly carried out for more general case with~\eqref{twistgen}. As
discussed before, the electroweak doublet leptons and down-type
quarks, that is $SU(5)$ anti quintuplet matter, have a redundancy of
flavor-dependent $E_6/G_{\rm SM}$ charges. Each twisting angle
$\theta_i$ sets the $\bar 5_i$ direction in $27$-plet and determines
its $U(1)_X\times U(1)_Z\subset E_6/SU(5)$ quantum numbers. As a
result, these scalar fields receive non-degenerate masses via the
$D$-term contribution and RG effects.

\medskip

Let us first consider the $D$-term contribution. It is known that the
$D$-term contribution arises associated with rank reduction of gauge
symmetry and its form depends on symmetry breaking pattern. There are
several patterns of the unified gauge symmetry
$E_6$~\cite{chain}. However in the present case of $E_6\to SU(5)$, 
the $D$-term contribution is generally parametrized by two mass
parameters without assuming explicit breaking chain (see 
Appendix \ref{sec:dterm}). For an eigenstate of $U(1)_X\times U(1)_Z$,
it is given by the following form:
\begin{eqnarray}
  \label{dterm1}
  \Delta m^2_i &=& (T_X)_i D_X + (T_Z)_i D_Z,
\end{eqnarray}
where $(T_X)_i$ and $(T_Z)_i$ denote the $U(1)_X$ and $U(1)_Z$ charges
of the field labeled $i$, and $D_{X,Z}$ are the mass parameters which
represent the deviation of the vacuum from supersymmetric flat
directions. It is noted that the above parametrization is also valid
in the case that $G_U$ is any subgroup of $E_6$ if one takes an
appropriate limit of $D$ terms.

The equation~\eqref{dterm1} is readily extended to the cases where
matter fields are not eigenstates of the broken $U(1)$ 
symmetries [i.e.\ $\theta_i\neq 0, \pi/2$ in~\eqref{twist1}]. The
$D$-term contributions to scalar masses of quark 
doublets $(m_{Q_i}^2)$, up-type quarks $(m_{\bar u_i}^2)$, down-type
quarks $(m_{\bar d_i}^2)$, lepton doublets $(m_{L_i}^2)$, and charged
leptons $(m_{\bar e_i}^2)$ are written as follows:
\begin{eqnarray}\notag
  \Delta m_{Q_i}^2 \;=\; \Delta m_{\bar u_i}^2
  \;=\; \Delta m_{\bar e_i}^2 &=& -D_X+D_Z, \\
  \label{dterm2}
  \Delta m_{L_i}^2 \;=\; \Delta m_{\bar d_i}^2 &=& x_iD_X+z_iD_Z,
\end{eqnarray}
where the coefficients $x_i$ and $z_i$ are 
\begin{eqnarray}
  x_i &=& 3\cos^2\theta_i-2\sin^2\theta_i, \\
  z_i &=& \cos^2\theta_i-2\sin^2\theta_i,
\end{eqnarray}
and $D_X$ and $D_Z$ are the mass parameters. It should be noted that
when one considers the cases where $SO(10)$ gauge symmetry is realized
at some intermediate stage of symmetry breaking chain, then $D_X$ and
$D_Z$ are directly connected to the $D$-term vacuum expectation values
of $U(1)_X$ and $U(1)_Z$.

Let us first focus on the particular limit $D_Z=0$, which is the case
that $E_6$ is broken down to $SO(10)$ or $SU(5)\times U(1)_X$ without
supersymmetry-breaking corrections. With the universal hypothesis for
soft terms at some high-energy scale, the induced scalar masses read
from \eqref{dterm2}
\begin{eqnarray}
  \notag
   m_{10}^2 &\equiv& m_{Q_i}^2 \;=\; m_{\bar u_i}^2 
   \;=\; m_{\bar e_i}^2 \;=\; m_0^2-D_X, \\
   m_{\bar 5_i}^2 &\equiv& m_{L_i}^2 \;=\; m_{\bar d_i}^2 
   \;= \;m_0^2 +(3\cos^2\theta_i-2\sin^2\theta_i)D_X,
\end{eqnarray}
at the GUT scale. From these mass expressions, we obtain a set of sum
rules among three-generation scalar masses:
\begin{eqnarray}
  \label{mass1}
  && \hspace{1.6cm}  
  \frac{m^{2\,(-)}_{ij}}{m^{2\,(+)}_{k\ell}} \;=\;
  \frac{\cos^2\theta_i-\cos^2\theta_j}{\cos^2\theta_k
    +\cos^2\theta_\ell-2/5}\,, \\[2mm] \notag
  && m^{2\,(-)}_{ij} \;=\; m_{\bar 5_i}^2-m_{\bar 5_j}^2\,, \qquad 
  m^{2\,(+)}_{k\ell} \;=\; m_{\bar 5_k}^2+m_{\bar 5_\ell}^2-2m_{10}^2\,.
\end{eqnarray}
The similar relations follow in the $D_X=0$ limit;
\begin{eqnarray}
  \label{mass2}
  \frac{m^{2\,(-)}_{ij}}{m^{2\,(+)}_{k\ell}} \;=\;
  \frac{\cos^2\theta_i-\cos^2\theta_j}{\cos^2\theta_k
    +\cos^2\theta_\ell-2}\,.
\end{eqnarray}
It is interesting to notice that these mass relations \eqref{mass1}
and \eqref{mass2} are parametrized only by the twist angles and 
independent of the mass parameters $m_0^2$ and $D$'s which are
controlled by the detail of supersymmetry-breaking dynamics in
high-energy regime. Therefore referring to these relations, one can
extract clear information on the generation structure in unified theory.

As illustrative examples, consider the above sum rules for the first
and second generation scalars, i.e., $(i,j)=(k,\ell)=(1,2)$. Since the
numerators of the formulae are proportional to
$\sin(\theta_1-\theta_2)$, the relative size of two twisting angles is
important. For $\theta_1=0$, which corresponds to the case that all
the first-generation matter fields originate from a single 16-plet 
of $SU(10)$, the scalar mass relations \eqref{mass1} and \eqref{mass2}
become 
\begin{eqnarray}
  \label{theta10}
  \frac{m^{2\,(-)}_{12}}{m^{2\,(+)}_{12}} \;=\;
  \left\{\begin{array}{cl}
      \dfrac{\sin^2\theta_2}{3/5+\cos^2\theta_2} 
      & \;\text{for} \;\, D_Z=0  \\[5mm]
      -1 & \;\text{for} \;\, D_X=0
    \end{array}\right.
\end{eqnarray}
On the other hand, for another limit $\theta_1=\pi/2$, the relations
take the forms
\begin{eqnarray}
  \frac{m^{2\,(-)}_{12}}{m^{2\,(+)}_{12}} \;=\;
  \left\{\begin{array}{cl}
      \dfrac{\cos^2\theta_2}{2/5-\cos^2\theta_2} 
      & \;\text{for} \;\, D_Z=0  \\[4mm]
      \dfrac{\cos^2\theta_2}{1+\sin^2\theta_2} 
      & \;\text{for} \;\, D_X=0
    \end{array}\right.
\end{eqnarray}
For the latter case ($D_X=0$) in \eqref{theta10}, the $\theta_2$
dependence drops out from the sum rule. This is because, due to the
absence of $SO(10)$-violating parameters $\theta_1$ and $D_X$, the
gauge symmetry ensures the degeneracy of the first-generation scalar
masses $m_{\bar 5_1}^2$ and $m_{10_1}^2$. Also a similar observation
appears in the case $D_Z=0$, where the $\theta_2$ dependence vanishes
for $\cos^2\theta_1=1/5$.

\medskip

There generally appear the corrections to scalar soft masses from RG
evolution above the unification scale. For the contributions from GUT
gauge interactions, scalar masses generally receive the gaugino
corrections according to their $E_6/G_{\rm SM}$ charges. Therefore the
non-degeneracy among different $\bar 5_i$ matter scalars which survive
to low-energy regime depends only on the differences among the
twisting angles:
\begin{eqnarray}\notag
  m_{\bar 5_i}^2 &=& m^2_{\bar 5\,(16)}\cos^2\theta_i
  +m^2_{\bar 5\,(10)}\sin^2\theta_i\,, \\
  \label{bcgen}
  &=& m_0^2+\tilde D\cos^2\theta_i\,,
\end{eqnarray}
where $m^2_{\bar 5\,(16)}$ [$m^2_{\bar 5\,(10)}$] is the
supersymmetry-breaking mass parameter for $\bar 5$ scalar fields which
belong to $16$-plets ($10$-plets) of $SO(10)$. It should be noted
that, even when the flavor-violating effects from $D$-term
contribution and RG evolution are included, we keep the 
form \eqref{bcgen} without the loss of generality, taking a suitable
redefinition of $m_0^2$ and $\tilde D$. As a result, a set of general
sum rules among the masses of $\bar 5$ scalar fields are obtained:
\begin{eqnarray}
  \label{massgen}
  && \frac{m^{2\,(-)}_{ij}}{m^2_{k\ell n}} \;=\; 
  \frac{\cos^2\theta_i-\cos^2\theta_j}{\cos^2\theta_k
    +\cos^2\theta_\ell-2\cos^2\theta_n}\,, \\[2mm]
  &&\hspace{7mm} \notag
  m^2_{k\ell n} \;=\; m_{\bar 5_k}^2+m_{\bar 5_\ell}^2-2m_{\bar 5_n}^2.
\end{eqnarray}
Note that the relation becomes trivial if some of the generation
indices take a common value in the denominator (or in the
numerator). That is, \eqref{massgen} is meaningful only 
when $k\neq\ell\neq n\neq k$, and therefore necessarily includes the
third generation matter. Like the 
relations \eqref{mass1} and \eqref{mass2}, this mass relation is also
independent of the high-energy free parameters $m_0$ and $\tilde D$,
and therefore useful to directly probe the flavor structure in
high-energy theory.

In the present example, the quark doublets, up-type quarks and charged
leptons have no twists and the induced spectrum is universal at the
boundary scale. If future particle experiments would observe some
deviations from such universality, that would provide an evidence that
the unified theory has non-trivial and more complicated matter
structure beyond the simplest one discussed in this paper.

\bigskip
\section{Probing flavor structure with low-energy observables}
\label{sec:low}
In this section, we perform the extraction of flavor structure in
unified theory through the twisting angles using the sum 
rules \eqref{mass1}, \eqref{mass2} and \eqref{massgen} among sfermion
soft masses. The scalar mass corrections are induced by the gauge
interactions of $G_U/G_{\rm SM}$ symmetry and have imprints of
high-energy flavor theory. The sum rules are useful in that they are
independent of the overall mass scale and the sizes of $D$-term
corrections. Therefore the generation structure is directly probed
from scalar mass parameters which would be expected to be in principle
revealed in future particle experiments.

We start from the discussion about RG effects on the first and second
generation scalar masses between the unification scale $M_G$ and the
electroweak scale $M_S$. For the first two generations, the
unification-scale soft masses at one-loop order are written in terms
of low-energy scalar masses and the universal gaugino mass parameter
$M_{1/2}$ as follows:
\begin{eqnarray}\label{rgsol}
  m_i^2(M_G) &=& m_i^2(M_S)+\xi_iM_{1/2}^2 +\eta_iS(M_S), \\[2mm] \notag
  \xi_i &=& \sum_n {2\over b_n}C_2^{n}(i)
  \left(\frac{g_n^4(M_S)}{g_G^4}-1\right) ,\\ \notag
  \eta_i &=& \frac{3}{5b_1}(T_Y)_i
  \left(\frac{g_G^2}{g_1^2(M_S)}-1\right).
\end{eqnarray}
Here the indices $n=1,2,3$ denote the SM gauge groups with the
coupling constants $g_n$, which meet to the universal value $g_G$ at
$M_G$. The gauge beta function coefficients are given by 
$b_n=(-3,1,33/5)$ for the MSSM, and $C_2^n(i)$ is the quadratic
Casimir invariant for the scalar labeled $i$. The mass parameter $S$
is the hypercharge $D$-term given by $S={\rm Tr}\,[(T_Y)_im_i^2]$.

It is noted that the high-scale soft masses \eqref{rgsol} are
described only by the observables at future experiments; the gauge
couplings, gaugino masses, and sfermion masses. For example, the
universal gaugino mass is obtained by converting the `observed' gluino
mass through the RG equations as $M_{1/2}\simeq M_3/2.5$ (or similarly
from the neutralino or chargino masses). The Higgs soft masses are
relevant to the electroweak symmetry breaking, which we do not focus
on in this paper. The low-energy value of $S$ could be extrapolated by
the observed up-quark and electron scalar masses in this case and we
simply set $S=0$ in the following analysis. Using the RG 
solution \eqref{rgsol}, the mass combinations which appear in the sum
rules \eqref{mass1} and \eqref{mass2} are written by
{\allowdisplaybreaks%
\begin{eqnarray}\notag
  m^{2\,(-)}_{12}
  &=& m_{L_1}^2(M_S)-m_{L_2}^2(M_S) \\
  &=& m_{\bar d_1}^2(M_S)-m_{\bar d_2}^2(M_S), \\ \notag
  m^{2\,(+)}_{12}
  &=& m_{L_1}^2(M_S)+m_{L_2}^2(M_S)-2m_{\bar e_1}^2(M_S)
  -2(\xi_L-\xi_e)M_{1/2}^2 \\
  &=& m_{\bar d_1}^2(M_S)+m_{\bar d_2}^2(M_S)-2m_{\bar e_1}^2(M_S)
  -2(\xi_d-\xi_e)M_{1/2}^2.
\end{eqnarray}}%
The coefficients of RG gauge contribution are evaluated by the
one-loop RG equations for gauge couplings;
$\xi_L\simeq 0.49$, $\xi_e\simeq 0.15$ and $\xi_d\simeq 4.35$
for $M_S\simeq 1$~TeV\@.

\begin{figure}[t]
\begin{center}
  \includegraphics[width=8cm,clip]{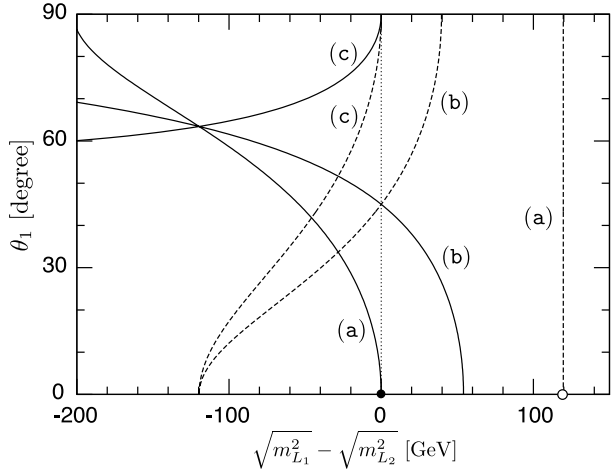}
\end{center}
\caption{The extracted values of twisting angles from the first and
second generation scalar lepton doublets. The solid (dashed) lines are
obtained from the formulae \eqref{mass1} [\eqref{mass2}], and the
lines (a), (b), (c) correspond to $\theta_2=0,\pi/4,\pi/2$,
respectively. The gaugino and scalar masses are fixed as
$M_{1/2}=300$~GeV, $m^2_{L_1}(M_S)+m_{L_2}^2(M_S)=2\times(500~{\rm GeV})^2$, 
and $m_{e_1}^2(M_S)=(400~{\rm GeV})^2$.\bigskip}
\label{fig:test}
\end{figure}

The twist angles for the first and second generations now become the
functions of mass parameters of scalar leptons (and down-type
quarks). In Fig.~\ref{fig:test}, we show using the 
formulae \eqref{mass1} and \eqref{mass2} the extracted twisting angles
from the low-energy non-degeneracy of scalar leptons.
For degenerate scalar leptons, the twist angles are identical
($\theta_1=\theta_2$). In general, a larger mass difference implies 
a larger deviation from the naive alignment
$\theta_1=\theta_2$. Notice however that there exist the points where
the two angles $\theta_1$ and $\theta_2$ are independent to each
other. For the case with \eqref{mass1} [\eqref{mass2}], that is given
by $\theta_1=\arccos(\frac{1}{\sqrt{5}})$ [$\theta_1=0$], at which
scalar masses of lepton doublets and charged leptons in the first
generation are common at the unification scale. Such a behavior is not
affected by the overall scale of superparticle masses. A
mass-dependent irregular line also appears if
$\sqrt{m^2_{L_1}}-\sqrt{m^2_{L_2}}\simeq120$~GeV, where
$\theta_2=\pi/2$ is predicted for any value of $\theta_1$. This is the
point that the masses of the second-generation scalar leptons are
unified.

Next let us examine the formula \eqref{massgen}. The third generation
scalars are necessarily incorporated in the analysis. They generally
receive non-negligible Yukawa effects through the RG evolution. The
size of RG Yukawa effects depend on $\tan\beta$ which is the ratio of
up- and down-type Higgs vacuum expectation values. For a sufficiently
small value of $\tan\beta$, the bottom and tau Yukawa couplings become
small, $y_{b,\tau}\ll 1$, and the RG contribution to scalar masses is
negligible. In this case, all the RG effects shown in \eqref{rgsol}
are canceled out in the mass formula \eqref{massgen}. In particular,
the gaugino masses are completely irrelevant to the analysis. We can
extract the twisting angles from low-energy masses of scalar leptons as
\begin{eqnarray}\label{massgen2}
  \frac{\cos^2\theta_1-\cos^2\theta_2}{\cos^2\theta_1+\cos^2\theta_2
    -2\cos^2\theta_3} &=&
  \frac{m_{L_1}^2(M_S)-m_{L_2}^2(M_S)}{m_{L_1}^2(M_S)+m_{L_2}^2(M_S)
    -2m_{L_3}^2(M_S)}\,.
\end{eqnarray}
A similar result follows for scalar down-type quarks.
\begin{figure}[t]
\begin{center}
  \includegraphics[width=6cm,clip]{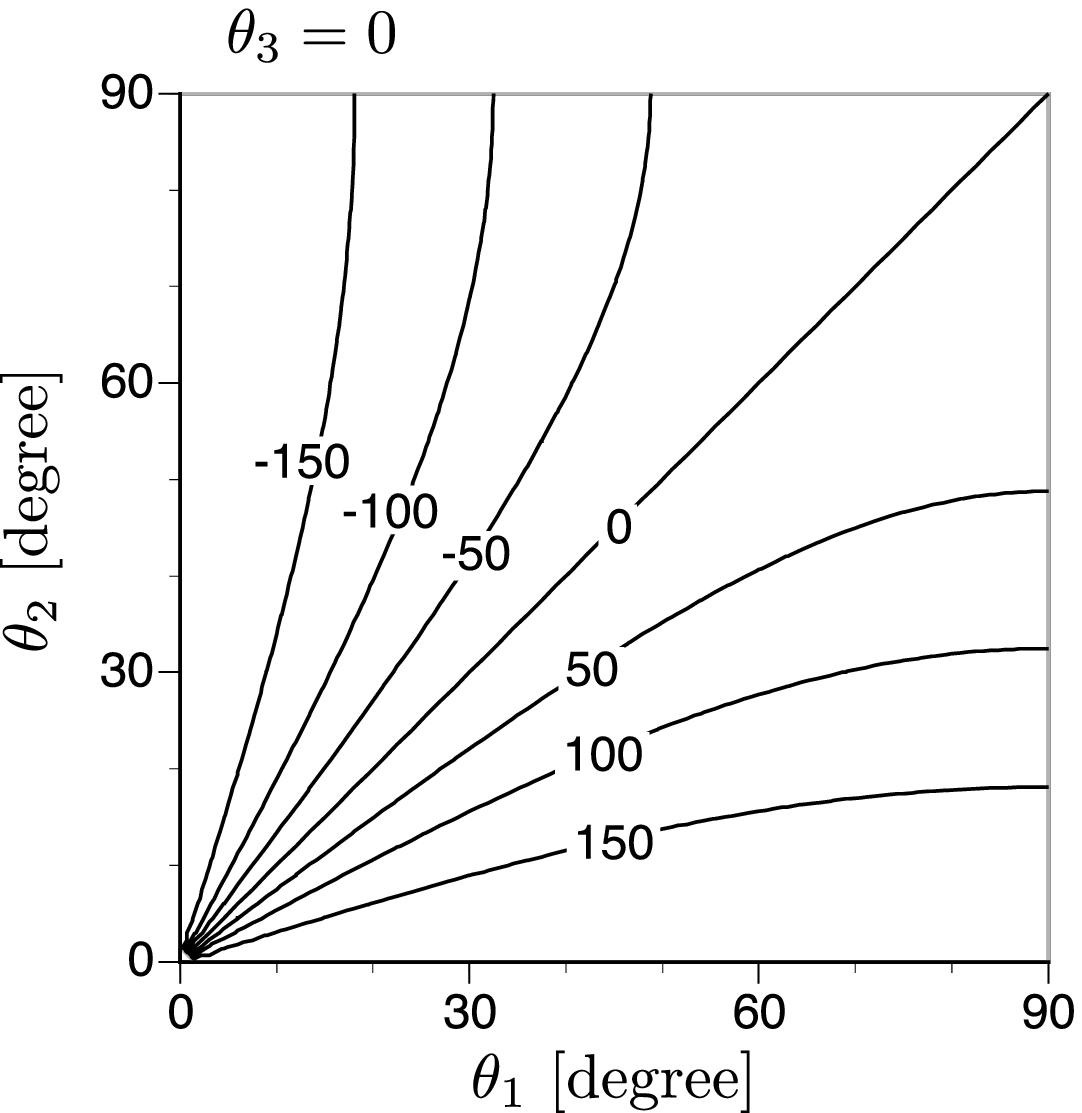}\hspace{1cm}
  \includegraphics[width=6cm,clip]{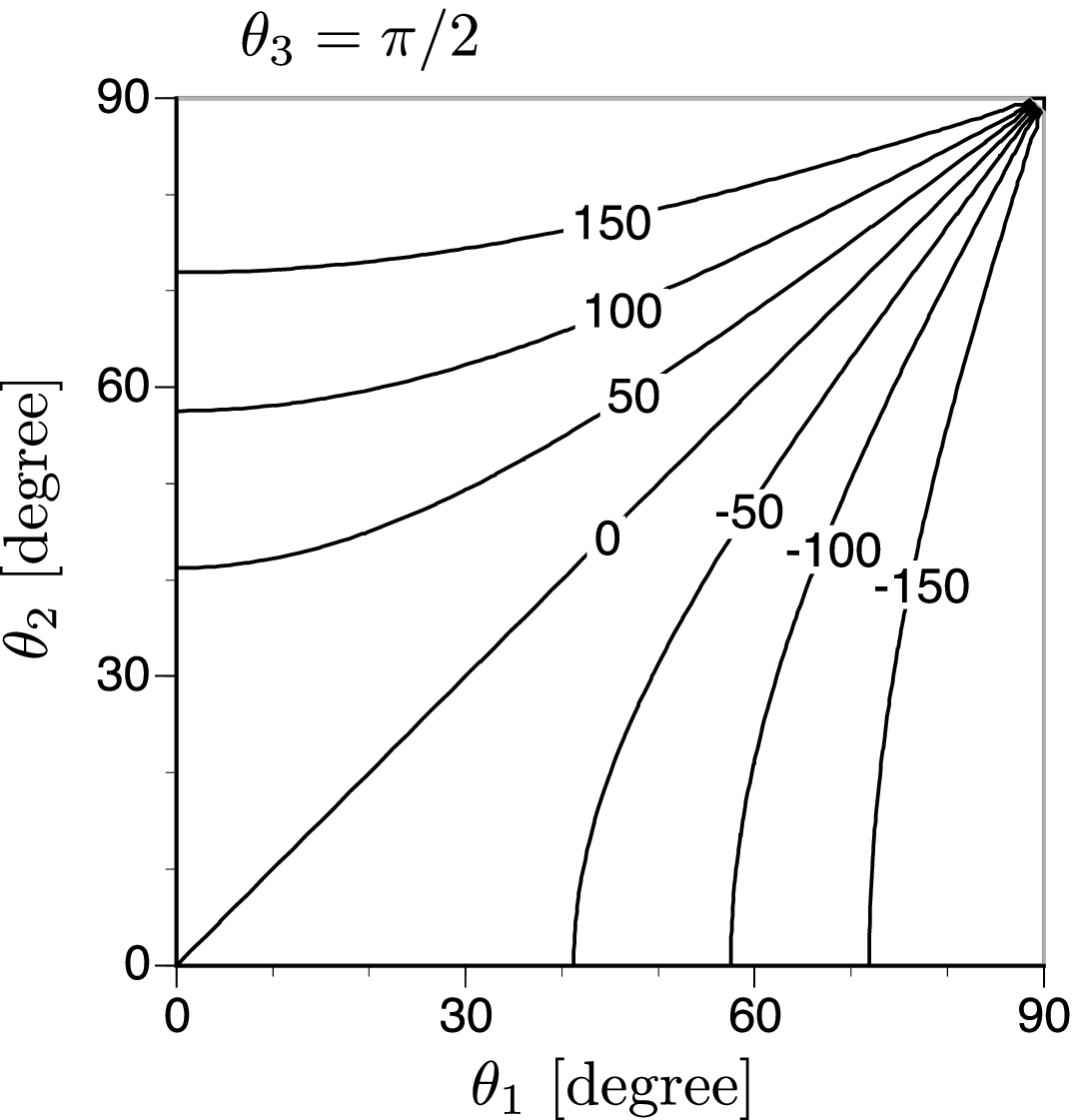}
\end{center}
\caption{The correlation between the twist angles $\theta_1$ and
$\theta_2$ indicated by the sum rule \eqref{massgen2}. The mass
difference $\sqrt{m_{L_1}^2}-\sqrt{m_{L_2}^2}$ is shown in the
figures. The scalar masses are fixed as 
$m^2_{L_1}(M_S)+m_{L_2}^2(M_S)=2\times (500~{\rm GeV})^2$ and
$m_{L_3}^2(M_S)=(400~{\rm GeV})^2$ as an example.\bigskip}
\label{fig:norge}
\end{figure}
Fig.~\ref{fig:norge} shows the correlation between the twisting angles
of the first and second generation scalar leptons. It is found from
the figure that the non-degenerate mass spectrum implies a deviation
from the naive alignment of twisting angles. The results are also
sensitive to the third-generation matter structure. As typical
examples, we show the cases that $\theta_3=0$ and $\pi/2$.

If $\tan\beta$ is not so small enough to neglect the Yukawa dependent
RG effects to the third generation scalar masses, one should evaluate
the contribution to the linear combination $m^2_{123}$. That is
carried out by solving the MSSM RG equations numerically.
\begin{figure}[t]
\begin{center}
  \includegraphics[width=8.4cm,clip]{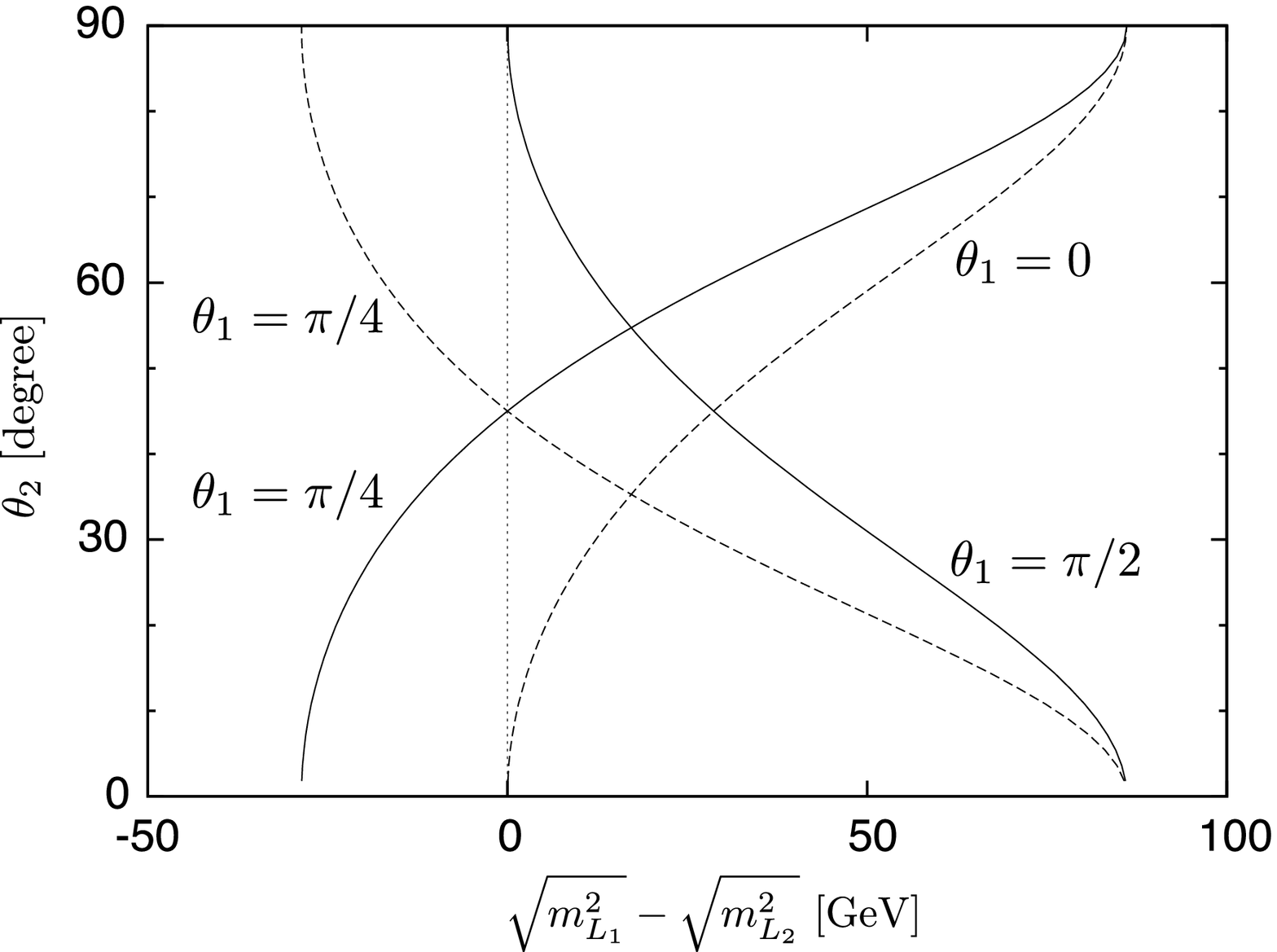}
\end{center}
\caption{A twisting angle as the function of the mass difference
between the first two generations scalar leptons. The solid (dashed)
lines represent the case that $\theta_3=0$ ($\theta_3=\pi/2$). We
assume $\tan\beta=10$ and $M_{1/2}=300$~GeV. The low-energy scalar
lepton masses are fixed as 
$m^2_{L_1}(M_S)+m_{L_2}^2(M_S)=2\times(700~{\rm GeV})^2$, 
$m_{e_1}^2(M_S)=(500~{\rm GeV})^2$, and 
$m_{L_3}^2(M_S)=(650~{\rm GeV})^2$.\bigskip}
\label{fig:wrge}
\end{figure}
In Fig.~\ref{fig:wrge}, we show the prediction of twisting angles
for $\tan\beta=10$. The scalar masses (difference) and the generation
angles are found to be tightly correlated even when the Yukawa RG
effects are taken into account.

In the present example, the relative size of non-degeneracy between
scalar leptons and down-type quarks is connected by the $SU(5)$
symmetry. Since the twist redundancy appears only for anti quintuplet
matter, the other MSSM matter scalars in the first and second
generations are expected to obtain the degenerate masses. Such a
significant pattern of non-universality is a particular prediction and
interesting signal of the $E_6$ unified theory with twisted matter
structure in a $27$-plet.

\medskip

If there were other sizable flavor-violating contributions to scalar
masses in unified theory (e.g.\ flavor-dependent RG effects), then the
mass relations are disturbed. Such contributions may be extracted
under model-dependent assumptions and is expected to be small, in
particular, for the first and second generations. Another possible
source of violation is a gauged flavor symmetry, often introduced in
order to explain the observed Yukawa coupling hierarchy. The
contributions from flavor gauge sector ($D$ term and gaugino
mass) \cite{flavorD} can also be treated in a similar manner to the
above analysis as long as the parametrization \eqref{twistgen}
holds. Further the $D$-term corrections which arise in the electroweak
symmetry breaking are not included in the analysis. This is due to the
fact that such effects are always canceled out in the combinations
$m_{ij}^{2\,(-)}$ and $m^2_{k\ell n}$, and for $m^{2\,(+)}_{12}$ the
contribution is generally small. These factors can be
straightforwardly incorporated into the analysis, if desired, at the
cost of complexity.

\medskip

In this section, we demonstrate that the generation structure in
unified theory is directly examined by using the information of
sfermion mass parameters which would be observed in future
experiments. The analysis can be generalized to other unified model
examples. The scalar mass spectrum in low-energy regime and their
various sum rules offer useful insights into un-revealed flavor
dynamics in high-energy fundamental theory.

\bigskip
\section{Flavor twists and flavor violation}
\label{sec:lfv}
In widely studied MSSM parameter space, for instance, with the
universal mass parameters at the unification scale, the scalar masses
of the first two generations are degenerate. Such a degeneracy is the
most naive and often adopted hypothesis for evading the experimental
constraints from flavor-violating processes~\cite{flavorvio} so that
non-SM sources of flavor violation do not exist at the unification
scale. Namely, off-diagonal elements of sfermion mass matrices vanish
in the super-CKM (SCKM) basis, which diagonalizes the
(scalar)-(fermion)-(gaugino) vertex for mass eigenstates of matter
fermions. In this case, the flavor violation triggered by
superparticles only appears via quantum corrections, which are often
small enough to avoid the experimental bounds.

With the generation redundancy of matter embedding \eqref{twistgen}, 
scalar fields generally realize non-degenerate mass spectrum. Such a
non-degeneracy does not ensure vanishing off-diagonal elements of
scalar mass matrices in the SCKM basis. In this section, we study the
experimental limitation on matter flavor structure in unified theory,
e.g.\ the twisting angles in the $E_6$ example \eqref{twist1}. The
constraints come from flavor-violating rare processes, in particular,
focusing on the charged lepton sector. A similar analysis can also be
performed with the general case \eqref{twistgen}.

To study the constraints on generation structure, one should fix the
relation between the two flavor bases: in one basis the Yukawa
interactions are diagonalized and in the other basis scalar soft mass
matrices are. We call the basis where scalar mass matrices have no
off-diagonal entries as ``the scalar basis''. In the following, we
treat the mass matrix for scalar lepton doublets, which matrix is
relevant for the flavor-violating rare processes in the lepton
sector. To turn to the SCKM basis from the scalar one, the scalar
lepton mass matrix is rotated as
\begin{eqnarray}
  m_{L_{ij}}^2 &=&
  \begin{pmatrix}
    m_{L_1}^2 & & \\
    & m_{L_2}^2 & \\
    & & m_{L_3}^2
  \end{pmatrix}
  \;\;\to\;\; m_{L_{ij}}^{2\,({\rm SCKM})} =\,
  V_{eL}
  \begin{pmatrix}
    m_{L_1}^2 & & \\
    & m_{L_2}^2 & \\
    & & m_{L_3}^2
  \end{pmatrix}V_{eL}^\dagger,
\end{eqnarray}
where $V_{eL}$ is defined by the charged-lepton Yukawa 
matrix $(Y_e)_{ij}$ in the scalar basis as 
\begin{eqnarray}
  Y_eY_e^\dagger &=& V_{eL}^\dagger 
  \begin{pmatrix}
    |y_e|^2 & & \\
    & |y_\mu|^2 & \\
    & & |y_\tau|^2
  \end{pmatrix}V_{eL}.
\end{eqnarray}
The lepton mixing matrix~\cite{MNS} is defined by
\begin{eqnarray}
  V_{\rm MNS} &=& V_{eL}V_{\nu}^\dagger,
\end{eqnarray}
where $V_{\nu}$ rotates the left-handed neutrinos such that the
neutrino mass matrix is diagonalized.

The recent neutrino experiments have revealed that the lepton mixing
matrix $V_{\rm MNS}$ has relatively large 1-2 and 2-3 mixing, and on
the other hand, the 1-3 mixing angle is found to be
small~\cite{neu}. These experimental results on the lepton mixing
matrix, however, does not directly restrict $V_{eL}$ alone as long as
the form of $V_{\nu}$ is undetermined. It is natural to imagine that
the observed large mixing angles have definite origins. In this case,
there are three possible patterns: (i) $V_e$ ($V_\nu$) has large 2-3
(1-2) angle, (ii) $V_e$ has both large mixing angles, and (iii)
$V_\nu$ has both large mixing angles. The last pattern may be
interesting in that the tiny neutrino mass generation simultaneously
enhances the neutrino mixing angles~\cite{enhance} while charged field
mixings remain small. In this section, however, we do not consider the
case (iii) since no significant requirement on twisting angles cannot
be obtained. Thus the experimental bounds on non-aligned twisting
angles depend on the form of charged lepton mixing $V_{eL}$.

Let us examine the first case that $V_e$ has large 2-3 mixing angle
(the large 1-2 mixing comes from the neutrino sector). We explicitly
consider the following form of $V_{eL}$:
\begin{eqnarray}
  V_{eL} &=&
  \begin{pmatrix}
    1 & \epsilon_x & \epsilon_z \\
    \frac{-\epsilon_x+\epsilon_z}{\sqrt{2}}
    & \frac{1}{\sqrt{2}} & \frac{-1}{\sqrt{2}} \\[1mm]
    \frac{-\epsilon_x-\epsilon_z}{\sqrt{2}} & \frac{1}{\sqrt{2}} &
    \frac{1}{\sqrt{2}}
  \end{pmatrix}, 
\end{eqnarray}
where the 2-3 generation mixing is maximal and the other ones
$\epsilon_x$ and $\epsilon_z$ are assumed to be small. The situation
is compatible with tiny 1-3 mixing without artificial fine tuning in
$V_{\rm MNS}$. With this form of $V_{eL}$, the scalar lepton mass
matrix in the SCKM basis is given by
\begin{eqnarray}
  m_L^{2\,({\rm SCKM})}\,\simeq\; 
\begin{pmatrix}
  m_{L_1}^2 &
  \frac{\epsilon_xm^{2\,(-)}_{21}-\epsilon_zm^{2\,(-)}_{31}}{\sqrt{2}} &
  \frac{\epsilon_xm^{2\,(-)}_{21}+\epsilon_zm^{2\,(-)}_{31}}{\sqrt{2}}
  \\[2mm]
  \frac{\epsilon_xm^{2\,(-)}_{21}-\epsilon_zm^{2\,(-)}_{31}}{\sqrt{2}} &
  \frac{m^{2\,(-)}_{21}+m^{2\,(-)}_{31}}{2} & 
  \frac{m^{2\,(-)}_{23}}{2} \\[2mm]
  \frac{\epsilon_xm^{2\,(-)}_{21}+\epsilon_zm^{2\,(-)}_{31}}{\sqrt{2}} &
  \frac{m^{2\,(-)}_{23}}{2} & \frac{m^{2\,(-)}_{21}+m^{2\,(-)}_{31}}{2}
\end{pmatrix}.
\end{eqnarray}
The off-diagonal elements of the mass matrix are relevant for lepton
flavor violating processes such 
as $\ell_j\to\ell_i\gamma$.\footnote{If neutrinos have sizable Yukawa
couplings, the off-diagonal elements in $m_L^2$ are radiatively
generated via RG evolution down to the decoupling scale of
right-handed neutrinos, and the induced lepton flavor violation has
been investigated in the literature~\cite{lfv}. For simplicity we do
not consider this issue in the present analysis.}
Among them, $\mu\to e\gamma$ tends to give the most severe constraints
on the flavor non-degeneracy, for which process the relevant
off-diagonal matrix element is
\begin{eqnarray}
  \label{offdiag}
  m_{L_{12}}^{2\,({\rm SCKM})} &=&
  \frac{\epsilon_xm^{2\,(-)}_{21}-\epsilon_zm^{2\,(-)}_{31}}{\sqrt{2}}.
\end{eqnarray}
A larger value of off-diagonal matrix elements increases the decay
amplitude. In other words, if $\epsilon_{x,z}$ are not suppressed, the
flavor non-universality $m^{2\,(-)}_{12}$ and $m^{2\,(-)}_{13}$, 
i.e.\ the twisting angles are strongly constrained unless a
cancellation occurs between the two terms in the right-handed side of
\eqref{offdiag}.

\begin{figure}[t]
\begin{center}
  \includegraphics[width=8.5cm,clip]{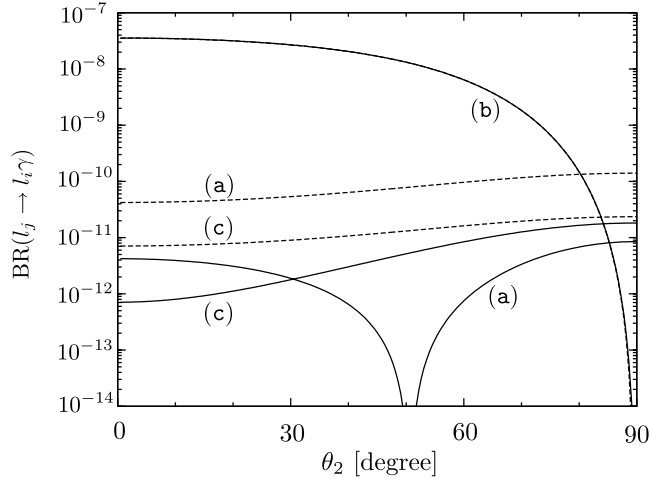}
\end{center}
\caption{Predictions of flavor-violating rare decay of charged leptons:
$\mu\to e\gamma$ [Line {\tt(a)}], $\tau\to \mu\gamma$ [Line {\tt(b)}]
and $\tau\to e\gamma$ [Line {\tt(c)}]. The solid lines show the case 
with $(\epsilon_x,\epsilon_z)=(10^{-2.3},10^{-2.5})$ and the dashed
ones with $(\epsilon_x,\epsilon_z)=(0,10^{-2})$. 
The $\tau\to\mu\gamma$ branching fraction is same for both cases. The
twist angles are set as $\theta_1=0$ and $\theta_3=\pi/2$ as examples
and superparticle masses are taken as in Fig.~\ref{fig:wrge}.\bigskip}
\label{fig:lfv}
\end{figure}

In Fig.~\ref{fig:lfv}, we show typical correlations between the 
$\ell_j\to\ell_i\gamma$ branching fractions and twisting angles.
The present experimental upper bounds on the fractions are 
${\rm BR}(\mu\to e\gamma)<1.2\times10^{-11}$,
${\rm BR}(\tau\to \mu\gamma)<6.8\times10^{-8}$ and 
${\rm BR}(\tau\to e\gamma)<1.1\times10^{-7}$ at the 90\% confidence 
level~\cite{PDG}. The validity of scenarios depends on the mixing
parameters $\epsilon_x$ and $\epsilon_z$. We here examine two typical
examples: the solid and dashed lines in the figure represent the cases
that $(\epsilon_x,\epsilon_z)=(10^{-2.3},10^{-2.5})$ and
$(0,10^{-2})$, respectively. At first, generally speaking, 
the $\mu\to e\gamma$ process gives the most stringent constraint on
the twisting angles. For the solid-line case, there exists a parameter
region where the cancellation between the two terms in \eqref{offdiag}
occurs. That decreases the $\mu\to e\gamma$ branching ratio and a
large generation twisting is generally allowed. For the dashed-line
case, the prediction is slightly larger than the experimental limit
and the scenario is marginally excluded. However in case that
$\epsilon_x$ is sufficiently small, the $\mu\to e\gamma$ amplitude
rather depends on $\epsilon_z$ and is 
suppressed as ${\rm BR}(\mu\to e\gamma)\lesssim 10^{-12}$
for $\epsilon_z\lesssim 10^{-3}$. That does not lead to any
experimental constraints. Secondly, the $\tau\to\mu\gamma$ process is
relevant to the non-degeneracy between the second and third
generations. Its amplitude is enhanced for $\theta_2\to 0$ where
the predicted branching ratio is just below the present experimental
bound and would be testable in near future experiments. 
The $\tau\to\mu\gamma$ decay is now induced through the large 2-3
generation mixing in $V_{eL}$ and therefore the constraint on matter
twist is insensitive to the mixing parameters $\epsilon_x$ and
$\epsilon_z$. Finally the $\tau\to e\gamma$ process strongly depends
on the details of $\epsilon_x$ and $\epsilon_z$. But we found that the
branching fraction is predicted to be much smaller than the
experimental bound and does not give any meaningful constraints on the
generation twisting as long as $\epsilon_z$ is small, as observed in
the neutrino experiments.

\begin{figure}[t]
\begin{center}
  \includegraphics[width=6.5cm,clip]{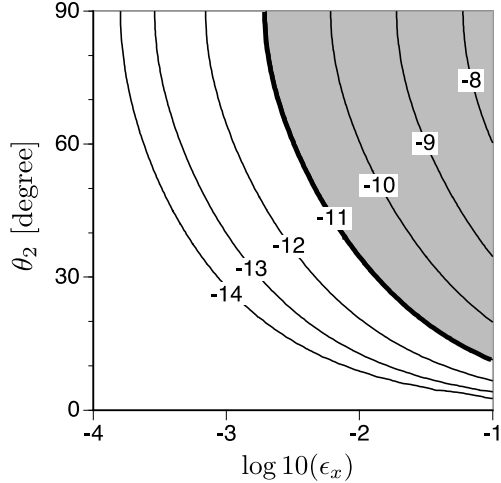}
\end{center}
\caption{Typical predictions for the $\mu\to e\gamma$ branching ratio.
The twisting angles are set to $\theta_1=0$ and $\theta_3=\pi/2$, and
$V_{eL}$ has the form \eqref{vel} with $\epsilon_y=\epsilon_z=10^{-2}$. 
The model parameters are similarly taken as in Fig.~\ref{fig:lfv}.
The gray shaded area is excluded by the current experimental upper
bound.\bigskip}
\label{fig:mueg}
\end{figure}

The second case we discuss is that $V_{eL}$ is the dominant origin for
both large lepton mixing angles. However, it is found in this case that
the scalar lepton non-degeneracy of the first two generations is
severely constrained by the experimental limit from 
the $\mu\to e\gamma$ rare process. As a result, the twisting angles
$\theta_1$ and $\theta_2$ should be almost aligned even if the other
mixing angles in $V_{eL}$ are sufficiently small. To see this
behavior, we show in Fig.~\ref{fig:mueg} the $\mu\to e\gamma$
branching ratio as a function of the 1-2 lepton mixing and twisting
angles. In the figure we take the explicit form of $V_{eL}$ as
\begin{eqnarray}
  V_{eL}&\simeq&
  \begin{pmatrix}
    1 & \epsilon_x & \epsilon_z \\
    -\epsilon_x & 1 & \epsilon_y \\
    -\epsilon_z & -\epsilon_y & 1
  \end{pmatrix},
  \label{vel}
\end{eqnarray}
with $\epsilon_y=\epsilon_z=10^{-2}$. That is, we have assumed a
conservative limit that the irrelevant lepton mixing angles are tiny
($\epsilon_{y,z}\ll1$), and then examine the $\epsilon_x$ dependence.
From the figure, one easily sees that the constraint from 
the $\mu\to e\gamma$ decay is rather severe for $\epsilon_x$, even
when the other mixing angles in $V_{eL}$ are small. Though the other
processes, $\tau\to \mu \gamma$ and $\tau\to e\gamma$, are sensitive
to $\epsilon_y$ and $\epsilon_z$, the amplitudes are not more than
those in Fig.~\ref{fig:lfv}.

The above discussion depends on supersymmetry-breaking soft mass
parameters. It is well known that the decay rates are enhanced for a
large value of $\tan\beta$~\cite{lfv}. Therefore for larger
$\tan\beta$, the constraints on the generation twisting become severer
than the cases in Figs.~\ref{fig:lfv} and \ref{fig:mueg}. The decay
rates also have dependences on superparticle mass scale; lighter
superparticles enhance the amplitudes.

In this section, we have examined the flavor-violating rare decays of
charged leptons in the presence of non-trivial generation structure in
high-energy unified theory. The results depend on the form of charged
lepton Yukawa matrix in the flavor basis where scalar lepton mass
matrices are diagonalized. If one generally considers other forms of
non-degeneracy \eqref{twistgen} and other flavor violation processes,
the analysis can be performed in completely parallel fashion.

\bigskip
\section{Summary}
\label{sec:sum}
In this paper, we have demonstrated that the flavor structure in
unified theory is probed with superparticle mass spectrum in
low-energy regime. The flavor dependence is parametrized by
the introduction of twisting angles which represent the embedding redundancy
of SM matter multiplets into unified theory. The non-universality is
generally induced in supersymmetry-breaking sfermion soft masses by
grand unified gauge interactions. We have obtained several sum rules 
for induced scalar masses: \eqref{mass1}, \eqref{mass2} 
and \eqref{massgen}. These relations are independent of explicit 
mass scales and are the functions of twist angles only. 
Therefore the relations are useful to directly probe the 
flavor sector in high-energy
regime.\footnote{For supersymmetry breaking parameters, 
a known scenario which could directly probe
high-energy property of the model is that the RG modification 
and supersymmetric threshold corrections are complemented by 
the contribution from conformal anomaly~\cite{M}.}

Using the mass relations, we have studied the twist angle behaviors
in the explicit $E_6$ example. It has been found that matter flavor
structure in unified theory is explored if future particle experiments
would reveal the superparticle mass parameters at low energy. The
physical implications from flavor-violating decay processes have also
been analyzed. The results depend on the form of Yukawa matrices in
the flavor basis where scalar lepton mass matrices are
diagonalized. It has been shown that non-aligned twist angles are
compatible with the experimental limits from lepton flavor violation
as long as there is no large off-diagonal elements of charged-lepton
Yukawa matrix between the first two generations.

The embedding redundancy naturally emerges if the fundamental unified
gauge theory is $SO(10)$ or larger, and/or with the presence of extra
matter. Future feedback from future particle experiments would reveal
specific patterns of sfermion mass spectrum, with which we would have
novel insights into the flavor structure in fundamental theory.

\bigskip
\subsection*{Acknowledgments}
This work is supported by grant-in-aid for the scientific research on
the priority area (\#441) "Progress in elementary particle physics of
the 21st century through discoveries of Higgs boson and supersymmetry"
(No.~16081209) and by a scientific grant from Monbusho
(No.~17740150).

\newpage
\appendix
\section*{Appendix}
\section{SM matter in $E_6$ multiplet}
\label{sec:27}
In this appendix, we summarize the SM matter assignment in unified
theory, in particular, the embedding into a $27$-plet of $E_6$ gauge
group. A main subject here is to show that the embedding of
right-handed down-type quarks and electroweak doublet leptons (and
also right-handed neutrinos) is not uniquely determined and can be
generation dependent, which fact leads to non-universal quantum
numbers of the broken gauge symmetry $E_6/G_{\rm SM}$.

In order to examine possible matter assignments, it is useful to 
consider the maximal subgroup decomposition of the vector
representation 27 under $E_6\supset SU(3)_C\times SU(3)_L\times SU(3)_R$:
\begin{eqnarray}
  27 &=& (3,3,1)\oplus (\bar 3,1,\bar 3)\oplus(1,\bar 3,3).
\end{eqnarray}
The first factor $SU(3)_C$ is identified with the $SU(3)$ color
group of the SM\@. Furthermore, under the fixing of electroweak
$SU(2)$ symmetry s.t.\ $SU(3)_L\supset SU(2)_L$, each representation
of $(SU(3)_C,SU(3)_L,SU(3)_R)$ is decomposed as
\begin{eqnarray*}
  (3,3,1) &=& (3,2,1)\oplus (3,1,1) \;\equiv\ (3,2)\oplus 3, \\
  (\bar 3,1,\bar 3) &=& (\bar 3,1,\bar 3)\;\equiv\,\bar 3^\alpha, \\
  (1,\bar 3,3) &=& (1,2,3)\oplus (1,1,3)\;\equiv\,2^\alpha \oplus 1^\alpha.
\end{eqnarray*}
Here we have used the notation that the labels of trivial
representation are dropped, and the superscript $\alpha$ represents
the $SU(3)_R$ index. Thus the components in a $27$-plet transform in
the following way under $E_6\supset SU(3)_C\times SU(2)_L$;
\begin{eqnarray}
  27 &=& (3,2)\,\oplus\, 3\,\oplus\, \bar 3^\alpha
  \,\oplus\, 2^\alpha \,\oplus\, 1^\alpha\,,
\end{eqnarray}
where $i=1,2,3$ denote the generation indices.

It is found that the $(3,2)$ multiplet is the unique choice for
electroweak doublet quarks in the SM\@. On the other hand, 
each $27$-plet contains three sets of $\bar 3$, $2$ and $1$, 
which correspond to $SU(3)_R$ triplets, and therefore the embedding of
other SM matter fields is not uniquely identified. The components of
$SU(3)_R$ triplets are discriminated with respect to the quantum
numbers of three $U(1)$ symmetries 
$[U(1)]^3\subset E_6/SU(3)_C\times SU(2)_L$. We take a basis for these
three $U(1)$ symmetries as
\begin{eqnarray}
  E_6 &\supset& SO(10)\times U(1)_Z \notag \\
  &\supset& SU(5)\times U(1)_X\times U(1)_Z \notag \\
  &\supset& SU(3)_C\times SU(2)_L \times U(1)_V\times U(1)_X
  \times U(1)_Z.
\end{eqnarray}
In a certain charge normalization, the $SU(3)_R$ singlets $(3,2)$ and
$3$ are found to have the following quantum numbers of
$(U(1)_V,U(1)_X,U(1)_Z)$ symmetry:
\begin{equation}
  Q((3,2)) = (1/6,-1,1), \qquad Q(3) = (-1/3,2,-2).
\end{equation}
With this normalization, the $SU(3)_R$ triplets $\bar 3^\alpha$, $2^\alpha$ 
and $1^\alpha$ take the following $U(1)$ charge assignment:
\begin{align}\notag
  Q(\bar 3^a)&=(-2/3,-1,1),& Q(2^a)&=(-1/2,3,1),& Q(1^a)&=(1,-1,1),\\
  \label{charges}
  Q(\bar 3^b)&=(1/3,3,1),& Q(2^b)&=(1/2,2,-2),& Q(1^b)&=(0,-5,1),\\
  Q(\bar 3^c)&=(1/3,-2,-2),& Q(2^c)&=(-1/2,-2,-2),& 
  Q(1^c)&=(0,0,4). \notag
\end{align}
(For the proper normalization of $E_6$ generators, these three $U(1)$
charges are multiplied by $\sqrt{3/5}$, $\sqrt{1/40}$, $\sqrt{1/24}$,
respectively). It is easily found from this assignment that the
$SU(3)_R$ triplets $\bar 3^\alpha$, $2^\alpha$ and $1^\alpha$ belong to the
multiplets of $SO(10)$ as
\begin{align*}
  16 &\,\ni\, \bar 3^a,\, \bar 3^b,\, 2^a,\, 1^a,\, 1^b, \\
  27 \,=\, 16\,\oplus\, 10\,\oplus\, 1, \qquad\qquad
  10 &\,\ni\, \bar 3^c,\, 2^b,\, 2^c, \\
  1 \,&\,\ni\, 1^c.
\end{align*}

The last factor undetermined is the hypercharge symmetry $U(1)_Y$ in
the SM\@. In general, the generator of $U(1)_Y$ is a linear
combination of the above three $U(1)$ generators. The successful gauge
coupling unification in the MSSM implies that the SM gauge group
$G_{\rm SM}=SU(3)_C\times SU(2)_L\times U(1)_Y$ is embedded into a
simple group at the unification scale. Under this situation, it is a
natural assumption that $E_6$ is broken down to a subgroup which
contains $SU(5)$ symmetry at a high-energy scale, and consequently the
hypercharge $U(1)_Y$ is identified with $U(1)_V$ in the above
decomposition of $E_6$ symmetry.

As a result, the residual SM matter fields, up-type quarks $\bar u_i$, 
down-type quarks $\bar d_i$, doublet leptons $ L_i$ and charged leptons 
$\bar e_i$, are found to be uniquely embedded into 27-plets as follows:
\begin{eqnarray*}
  \bar u_i &=& \bar 3^a_i, \\
  \bar d_i &=& \bar 3^b_i\cos\theta_i +\bar 3^c_i\sin\theta_i, \\
  L_i &=& 2^a_i\cos\theta_i +2^c_i\sin\theta_i, \\
  \bar e_i &=& 1^a_i.
\end{eqnarray*}
(The right-handed neutrinos are identified with a linear combination
$1^b_i\cos\theta_i+1^c_i\sin\theta_i$.) \ At this stage, there appear
three degrees of freedom $\theta_i$ ($i=1,2,3$), i.e.\ the ways of
embedding $\bar 5_i$ matter in the $SU(5)$ language, which we refer to
as the twisting angles in the text.

Finally we comment on the $B-L$ (baryon minus lepton number)
symmetry. If one assumes that $U(1)_{B-L}$ is a gauge symmetry
embedded in $E_6$ group, that is possible only when the matter
twisting angles take the boundary values $\theta_i=0$ or $\pi/2$. For
$\theta_i=0$, the generator of $U_{B-L}$ corresponds to
\begin{eqnarray}
  T_{B-L} &\propto& 4T_V-T_X,
\end{eqnarray}
and for $\theta_i=\pi/2$
\begin{eqnarray}
  T_{B-L} &\propto& 16T_V+T_X+5T_Z, 
\end{eqnarray}
where $T_V$, $T_X$ and $T_Z$ are the generators of $U(1)_V,U(1)_X$ and 
$U(1)_Z$. Therefore for the general matter embedding
($\theta_i\neq\theta_j$), the $B-L$ symmetry is not local one. The
continuous values of twisting angles implies that potential terms of
the model should possess abelian global symmetries. One linear
combination of these global and local $U(1)$ symmetries remains
intact in low-energy regime as $U(1)_{B-L}$ (which is accidentally
enhanced within the SM matter content).

\bigskip
\section{Multiple $\boldsymbol{D}$-term corrections to scalar masses}
\label{sec:dterm}
Here we present the general parametrization of $D$-term corrections to
scalar soft masses, which arise in the rank reduction of gauge
symmetry~\cite{D}. As an example, let us examine the case that $E_6$
is broken down to $SU(5)$ as the following symmetry breaking chain:
\begin{eqnarray}
  E_6 \,\to\, SU(6)\times SU(2) \,\xrightarrow{\Lambda_1}\,
  SU(5)\times U(1) \,\xrightarrow{\Lambda_2}\, SU(5),
\end{eqnarray}
where the breakdown occurs at two subsequent rank reduction scales 
$\Lambda_1$ and $\Lambda_2$. At each scale, there appears $D$-term
contribution to scalar masses. For the first reduction at $\Lambda_1$,
the correction is written as
\begin{eqnarray}
  \Delta_1 m_i^2 &=& [g_6^2(T_6)_i\cos\alpha+g_2^2(T_2)_i\sin\alpha]
  D_{\Lambda_1},
\end{eqnarray}
where $g_2$ and $g_6$ are the $SU(2)$ and $SU(6)$ gauge couplings,
respectively, and $D_{\Lambda_1}$ is a mass parameter (a deviation of
the symmetry-broken vacuum from supersymmetric directions). The
diagonal generators for vector representation are given by 
$T_6\propto {\rm diag}(1,1,1,1,1,-5)\otimes 1$ 
and $T_2\propto 1\otimes(1,-1)$, which commute with the $SU(5)$
generators. The unbroken $U(1)$ generator at $\Lambda_1$ is
parametrized by the angle $\alpha$ as $T_2\cos\alpha-T_6\sin\alpha$,
and the associated gauge coupling constant is given by
\begin{eqnarray}
  \frac{1}{g_1^2} &=& \frac{\cos^2\alpha}{g_2^2}+
  \frac{\sin^2\alpha}{g_6^2}.
\end{eqnarray}
At the second reduction scale $\Lambda_2$, there appears another
$D$-term contribution due to the breaking of the above $U(1)$ symmetry:
\begin{eqnarray}
  \Delta_2 m_i^2 &=& [g_1^2(T_2)_i\cos\alpha-g_1^2(T_6)_i\sin\alpha]
  D_{\Lambda_2}.
\end{eqnarray}
In the end, the total $D$-term correction is written as follows:
\begin{eqnarray}\notag
  \Delta m^2_i &=& \Delta_1 m^2_i+\Delta_2 m^2_i \\ \notag
  &=& [g_6^2(T_6)_i\cos\alpha+g_2^2(T_2)_i\sin\alpha]D_{\Lambda_1}
  +[g_1^2(T_2)_i\cos\alpha-g_1^2(T_6)_i\sin\alpha]D_{\Lambda_2}, \\
  &=& (T_6)_iD_6+(T_2)_iD_2,
\end{eqnarray}
where $D_{2,6}$ are the redefined mass parameters. An important fact
here is that the generators $T_6$ and $T_2$ can be rewritten by
suitable linear combinations of the generators 
of $U(1)^2\subset E_6/SU(5)$. Therefore if one chooses $U(1)_X$ and
$U(1)_Z$ as the basis of $U(1)^2$, the scalar mass correction is
generically parametrized in the following form:
\begin{eqnarray}
  \Delta m^2_i &=& (T_X)_i D_X + (T_Z)_i D_Z.
\end{eqnarray}
The parametrization is always valid for $D$-term contributions
associated with $E_6\to SU(5)$ breaking, since any $U(1)$ generator
in $E_6/G_{\rm SM}$ is spanned by $T_X$ and $T_V$. Thus scalar masses
generally have a wide variety due to generation-dependent charges
$(T_X)_i$ and $(T_Z)_i$, according to the matter embedding discussed
in Appendix \ref{sec:27}.

The result is easily generalized to the case with an arbitrary gauge
group $G_U$ which is a rank $n$ semi-simple group. Above the
unification scale, $G_U$ is broken down to $SU(5)$ and the $D$-term
contributions generally appear. One is always able to take a linearly
independent set of $U(1)$ generators $T_\ell$ ($\ell=1,\cdots, n-4$)
as the basis of Cartan generators of $G_U/SU(5)$. Using the generators
$T_\ell$, the general form of scalar mass corrections is written as
follows:
\begin{eqnarray}
  \Delta m^2_i &=& \sum_{\ell =1}^{n-4}(T_\ell)_i D_{\ell}.
\end{eqnarray}
The parametrization is valid for arbitrary symmetry breaking
chains. If the charges $(T_\ell)_i$ are generation dependent according
to the matter assignment in unified theory, generation-dependent
contributions to scalar masses arise from the $D$ terms.


\end{document}